# Logical characterizations of computational complexity classes

Vladimir Naidenko

Institute of Mathematics
National Academy of Sciences of Belarus
ul. Surganova 11, Minsk 220012, Belarus

**Abstract**

Descriptive complexity theory is an important area in the study of computational complexity. In this direction, it is possible to describe combinatorial problems exclusively by logical methods, without resorting to the use of complicated algorithms. The first work in this direction was written in 1974 by the American mathematician Fagin [1]. The article describes the development of methods of the theory of descriptive complexity.

Since 1974, descriptive complexity theory has been actively developed, in which computational complexity is characterized in terms of logical languages. In the theory of computational complexity, the set of all practically significant problems is divided into complexity classes that combine problems that are similar in terms of computational volume. Logical characterizations of complexity classes occupy an important place in computational complexity theory. So, if a class has its logical characterization, it means that the formal description of the problem can easily determine its computational complexity. Fagin [1] first showed that the complexity class coincides with the set of problems described by second-order existential logic. Stockmeyer [2] extended Fagin's result to a polynomial hierarchy , describing it using second-order logic. Further research revealed logical characterizations for several more complexity classes [3].

Complexity classes can be divided into syntactic and semantic. Syntactic classes are uniquely defined by algorithms with pre-defined limits on computing resources. At the same time, additional (semantic) conditions are imposed on semantic classes. For example, the famous NP complexity class includes the vast majority of applied computational problems (in the recognition variant with possible positive and negative YES and NO answers) solved by nondeterministic Turing machines. Moreover, the time (number of clock cycles) of such a machine depends on the size of the problem to be solved. At an intuitive level, the NP class consists of recognition problems for which it is easy (i.e., in a time limited by a polynomial of the problem size) to check whether the proposed solution is correct. The NP class is a syntactic class, since it is uniquely defined by a nondeterministic Turing machine. The coNP class consists of counterexamples of NP problems, where the counterexample of the problem has a positive solution (YES) if and only if the problem has a negative solution (NO). This class is also syntactic. The intersection of the NP∩coNP complexity classes NP and coNP occupies an important position in computational complexity theory. This class is already semantic because of the additional intersection condition. This class, in particular, plays a crucial role in public key cryptography [4], since the latter is largely based on the factorization problem underlying NP∩coNP.

A number of prominent experts in the field of logic and complexity theory (Dawar [5], Papadimitriou [6]) noted difficulties in finding logical characterizations for semantic complexity classes. To overcome these difficulties, we developed a universal approach to creating logical characterizations based on characteristic sets, the research of which was initiated in [7]. Using this approach, we obtain the following result, which represents a significant advance in the development of complexity theory: for the first time, we construct a recursive set of second-order existential logic formulas that precisely defines the complexity class NP∩coNP of combinatorial problems on arbitrary structures. This refutes the well-known hypothesis about the impossibility of recursive representation (or algorithmic enumeration) of all problems from the class NP∩coNP. Research on recursive representations of complexity classes has actually been conducted since 1974, but so far it has not been possible to find any recursive representation of NP∩coNP – one of the most important complexity classes. This has led many leading experts in the field of complexity theory and logic to consider it impossible to enumerate problems in the class NP∩coNP. Thus, one of the founders of modern computational theory, Professor Papadimitriou of Stanford University [6] emphasized the extreme difficulty of obtaining a recursive enumeration of problems from the class NP∩coNP. President of the European Association for logic in computer science Dawar [5] in 2010 at the International conference "Fields of Logic and Computation" (Fields of Logic and Computation) expressed the opinion that complexity classes defined by semantic restrictions on certifying machines, such as, for example, the classes NP∩coNP and RP, do not allow obvious recursive representations. Moreover, he believed that it is possible to find a recursive representation for the class NP∩coNP will require a fundamentally new characterization of the class and will be a major breakthrough in complexity theory: "Thus, finding a recursively enumerable set of witnesses would require a fundamentally new characterization of the class and would be a major breakthrough in complexity theory". The above result just gives a recursive representation of all problems from NP∩coNP. Moreover, on the basis of this representation, he first constructed a logic for the class NP∩coNP, which Dawar rated as "doubly improbable" (doubly unlikely) in the same report [5] at the conference.

We introduced [7] a logic for the NP∩coNP complexity class that makes it possible for the first time to describe all problems from NP∩coNP, which has not only fundamental, but also important applied knowledge in the field of information security, taking into account the central role of the NP∩coNP class in public key cryptography. Back in the 70s, [4] theoretically justified the direct dependence of the cryptographic strength of encryption on the computational complexity of problems from the NP∩coNP class. But until today, very little was known about the problems themselves from NP∩coNP. In fact, only two problems from this class were known (excluding problems from class P), namely, the problem of factorization of numbers and the problem of finding a sufficiently short vector of an integer lattice. For comparison, the class of NP-complete problems contains thousands of practically important problems and their number is constantly growing. If it suddenly turns out that the factorization problem has an effective solution (which is very likely in the light of recent success in solving the related problem of number simplicity), then all modern cryptography based on factorization will cease to exist. To avoid this, it is important to be able to find other problems in the NP∩coNP class as a potential replacement for the factorization problem. But how do I do this? If we use the classical (i.e., based on a nondeterministic Turing machine) definition of the class NP∩coNP, accepted in modern complexity theory, then determining whether any suitable candidate problem belongs to the class NP∩coNP leads to the need to solve the algorithmically unsolvable problem of algorithm equivalence. It was the presence of such an algorithmically unsolvable problem (in the classical approach to the class NP∩coNP) that served as the very argument that Dawar used to justify the impossibility of recursive representation of the class NP∩coNP. Therefore, it is not surprising that only two problems in the NP∩coNP class have been identified so far. Our approach avoids the difficulties associated with dealing with algorithmically unsolvable problems. Moreover, it makes it possible (at least theoretically) to efficiently list problems in the NP∩coNP class and choose from them suitable for use in cryptography.